\documentclass[%
 apl,%
 amsmath,%
 amssymb,%
 amsfont,%
 reprint,%
]{revtex4-1}

\usepackage{graphicx}
\usepackage{bm}

\begin{document}

\title{ Tuning lattice thermal conductance by porosity control \\ in ultra-scaled Si and Ge nanowires }%
\author{Abhijeet Paul}
\email{abhijeet.rama@gmail.com}
\author{Gerhard Klimeck}

\affiliation{School of Electrical and Computer Engineering, Network for Computational Nanotechnology, Purdue University,%
 West Lafayette, Indiana, USA, 47907.}

\date{\today}

\begin{abstract}

Porous nanowires (NWs) with tunable thermal conductance are examined as a candidate for thermoelectric (TE) devices with high efficiency (ZT). 
Thermal conductance of porous Si and Ge NWs is calculated using the complete phonon dispersion obtained from a modified valence force field (MVFF) model. The presence of holes in the wires break the crystal symmetry which leads to the reduction in ballistic thermal conductance ($\sigma_{l}$). $[$100$]$ Si and Ge NWs show similar percentage reduction in $\sigma_{l}$ for the same amount of porosity. A 4nm $\times$ 4nm Si (Ge) NW shows  $\sim$ 30\% (29\%) reduction in $\sigma_{l}$ for a hole of radius 0.8nm. The model predicts an anisotropic reduction in $\sigma_{l}$ in SiNWs, with $[$111$]$ showing the maximum reduction followed by $[$100$]$ and $[$110$]$ for a similar hole radius. The reduction in $\sigma_{l}$ is attributed to phonon localization and anisotropic mode reduction.


\end{abstract}


\maketitle 
\section{Introduction} \label{sec:I}

Extreme geometrical confinement and surfaces make nanowires (NWs) a promising candidate to obtain higher ZT thermoelectric material by suppressing the lattice thermal conductivity ($\kappa_{l}$). Recent experimental works \cite{Hochbaum2008,Boukai2008} reveal that nanowire geometry can greatly enhance the ZT of Si  (upto 1 at 200K \cite{Boukai2008}) from its bulk value of 0.06 at 300K \cite{TTvo_sinw}. An interesting way to further reduce the thermal conductivity in bulk Si was shown both experimentally \cite{expt_porous_si} and theoretically \cite{theory_porous_si} by creating pores in the crystalline material. More recently the fabrication of phonon `nano-mesh' \cite{expt_porous_si} showed a 100 fold reduction in $\kappa_{l}$ from the bulk Si value of 148 W/m-K to 1.9W/m-K. Recent technological developments allow to fabricate hollow nanowire arrays using sacrificial templates \cite{hollow_wire_sacrificial}, hollow spinel wires using the `Kirkendall effect' \cite{hollow_wire_kirkendall}, electrochemical anodic dissolution \cite{electro_deposition}, template based hollow wire fabrication \cite{template_hollow_wires}, etc. Thus theoretical investigation can guide these improved fabrication methods to obtain better TE materials and structures.


The stronger scattering of the phonons from the surfaces compared to the electrons, reduces $\kappa_{l}$ drastically without affecting the electrical conductivity (G) in porous materials due to the much smaller coherent phonon wavelengths than electron \cite{theory_porous_si}. Motivated by the interesting experimental works and theoretical arguments, we analyze the ballistic thermal conductance ($\sigma_{l}$) in hollow Si and Ge nanowires with various channel orientations ($[$100$]$, $[$110$]$ and $[$111$]$) to explore the candidates which lead to higher reduction in $\sigma_{l}$ and to understand the reasons for such reduction.

\begin{figure}[t]
	\centering
		\includegraphics[width=2.6in,height=2.5in]{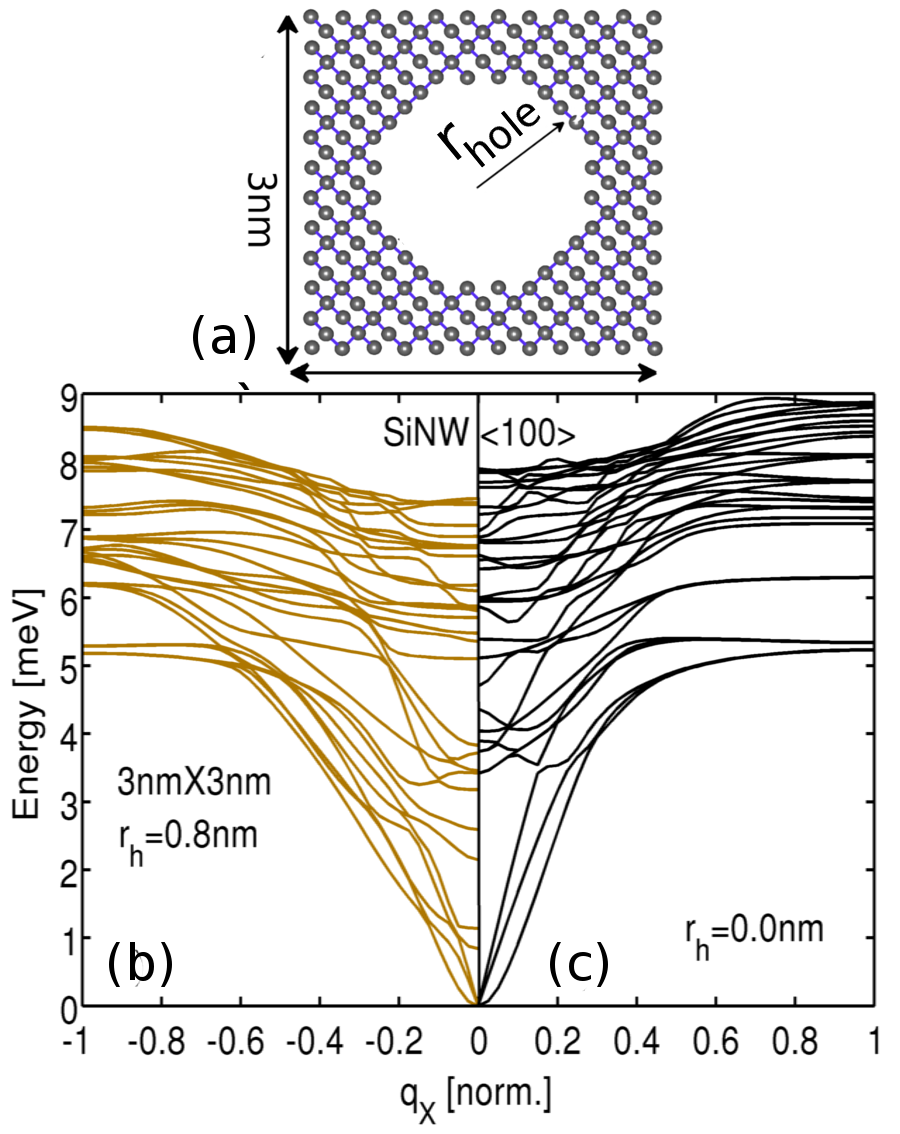}
	\caption{(a) Projected hollow Si unitcell with $\langle$100$\rangle$ channel orientation. Phonon dispersion obtained from the MVFF model for $\langle$100$\rangle$ (b) hollow and (c) filled SiNW with cross-section size of 3nm(W) $\times$ 3nm(H).}
	\label{fig:sinw_hollow}
\end{figure}

The following section (Sec. \ref{sec:II}) briefly discusses the approach for the calculation of the phonon dispersion and $\sigma_{l}$ in hollow Si and Ge NWs. The effect of the hole radius on $\sigma_{l}$ and the reasons for the reduction of $\sigma_{l}$ are discussed in Sec. \ref{sec:III}, followed by the conclusions (Sec. \ref{sec:conc}).

\section{Theory and Approach} 
\label{sec:II}

The phonon dispersion in free-standing Si and Ge nanowires is calculated using a frozen phonon approach, called the modified valence force (MVFF) model \cite{VFF_mod_herman,vff_own_paper,vff_iwce_paper} (Fig.\ref{fig:sinw_hollow} a). In the MVFF method the frequencies of selected phonon modes are calculated from the forces acting on atoms produced by finite and periodic displacements of the atoms in an otherwise perfect crystal at equilibrium. The MVFF model has successfully explained the bulk phonon dispersions as well as other lattice properties successfully in bulk Si, Ge, etc. \cite{VFF_mod_herman,vff_own_paper} and has recently been applied to SiNWs to explain their phonon spectra and lattice thermal properties \cite{vff_iwce_paper}. 

\begin{figure}[!b]
	\centering
		\includegraphics[width=3.3in,height=3.0in]{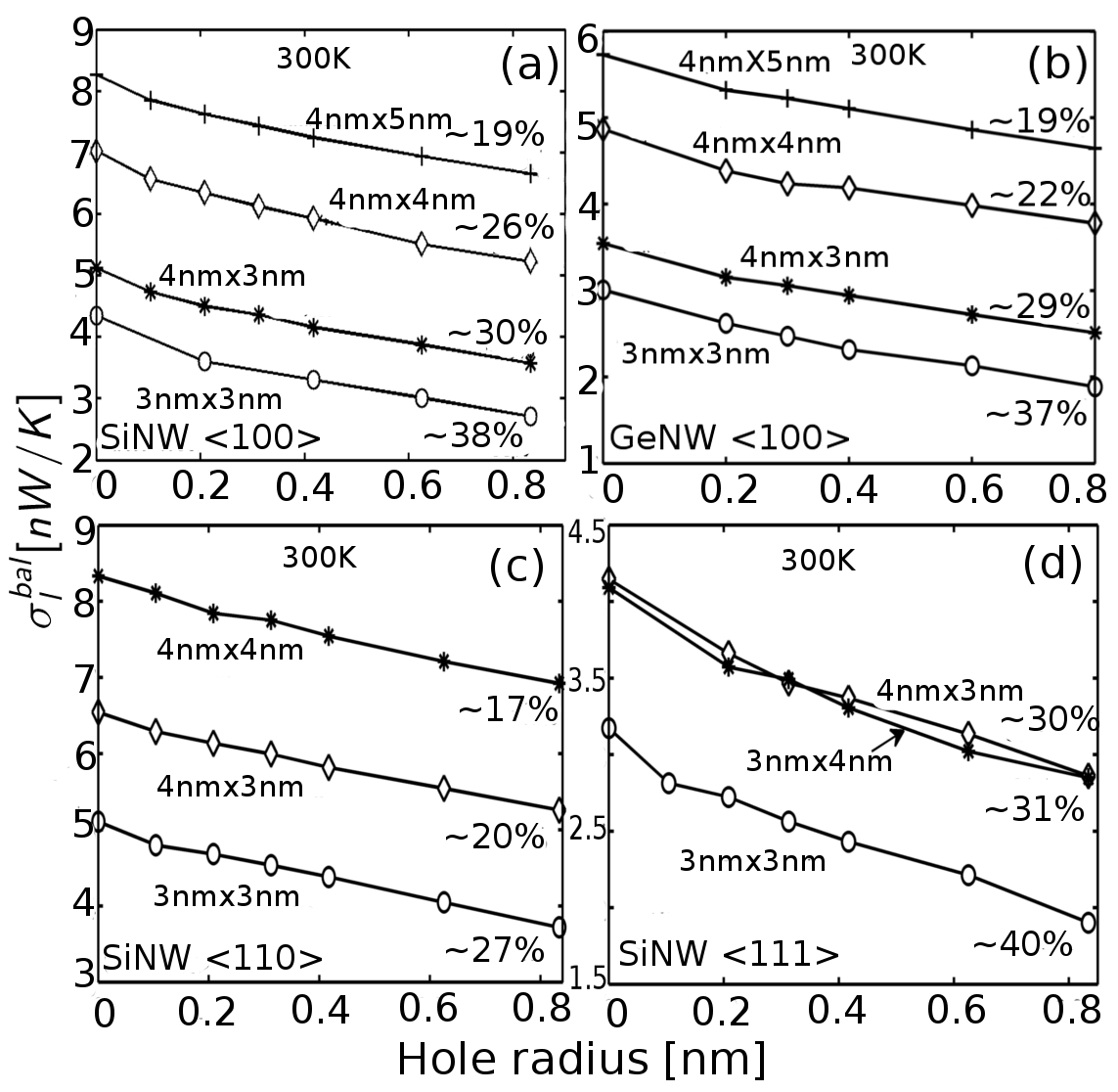}
	\caption{ Ballistic $\sigma_{l}$ in small cross-section rectangular nanowires. (a) $[$100$]$ SiNW, (b) $[$100$]$ GeNW, (c) $[$110$]$ SiNW and (d) $[$111$]$ SiNW. The percentage reduction in $\sigma_{l}$ for all the wires for $r_{h}$ = 0.8nm is also indicated.}
	\label{fig:nw_kthermal}
\end{figure} 

From the calculated phonon dispersion, the ballistic thermal conductance ($\sigma_{l}$) across a semiconductor slab/wire can be calculated using the Landauer's formula \cite{Land} using Eq.(\ref{eq_sigma}) \cite{jauho_method,mingo_ph},
\begin{footnotesize}
\begin{eqnarray}
\label{eq_sigma}
 	\sigma_{l}(T)& = &\frac{e^2}{\hbar} \int^{E_{max}}_{0} M(E)\cdot E \cdot \frac{\partial}{\partial T}\Big[\frac{1}{(exp\big(\frac{E}{k_{B}T})-1\big)} \Big],
\end{eqnarray}
\end{footnotesize}

where M(E), $k_{B}$,  $\hbar$ and  $e$ is the number of modes at energy E, the Boltzmann's constant, the Planck's constant and the electronic charge, respectively. This equation is valid when the two contacts across the material slab are maintained under a small temperature gradient $\Delta T$ \cite{mingo_ph}.

\textit{Device details:} Rectangular NWs are studied with width (W) and height (H) varying from 3nm to 4nm with three channel orientations of $[$100$]$, $[$110$]$ and $[$111$]$. The hole radius ($r_{h}$) in these NWs have been chosen such that not more than 25\% of the total atoms (Fig. \ref{fig:sinw_hollow} a) have been removed from the unitcell to ensure structural stability of these wires (no negative phonon dispersion is obtained \cite{SINW_110_phonon}). The inner and outer surface atoms in these NWs are allowed to vibrate freely (Fig. \ref{fig:sinw_hollow} a). In extremely small SiNWs ($W\le1nm$) significant atomic reconstruction can take place which has not been considered in this study \cite{Amrit}.  

\section{Results and Discussion}
\label{sec:III}

\textit{Phonon spectra:} Hollow nanowires have fewer atoms per unitcell compared to the solid nanowires which results in a reduced number of phonon sub-bands. Figure \ref{fig:sinw_hollow} (b) and (c) compare the phonon spectra of a 3nm $\times$ 3nm $[$100$]$ SiNW with $r_{h}$ = 0.8nm and 0nm, respectively. In the hollow NW a lot of sub-branches appear in the lower portion of the phonon spectra (Fig. \ref{fig:sinw_hollow} b) indicating more surface phonon confinement due to an increased surface to volume ratio (SVR).
   
\textit{Ballistic thermal conductance $\sigma_{l}$:} In all the nanowires $\sigma_{l}$ is calculated using Eq.(\ref{eq_sigma}). As a general trend, $\sigma_{l}$ reduces in hollow nanowires compared to the solid nanowires (Fig. \ref{fig:nw_kthermal}). Comparison of $\sigma_{l}$  in Si and Ge NW show a similar amount of reduction in thermal conductance (Fig. \ref{fig:nw_kthermal} a and b). This indicates a  weak material dependence of the reduction mechanisms. A 3nm $\times$ 3nm Si(Ge) NW show a reduction of $\sim$37\%(38\%) for $r_{h}$ = 0.8nm. This reduction decreases as the wire cross-section size increases due to the reducing SVR. The phonons feel the surface less in larger cross-section NWs compared to the smaller wires.

The $\sigma_{l}$ reduction is wire orientation dependent as revealed in Fig. \ref{fig:nw_kthermal} a, c and d. For a 3nm $\times$ 3nm SiNW with $r_{h}$ = 0.8nm, a $[$100$]$ wire shows a reduction of $\sim$38\% (Fig. \ref{fig:nw_kthermal} a), a $[$110$]$ wire shows a reduction of $\sim$27\% (Fig. \ref{fig:nw_kthermal} c) and a $[$111$]$ wire shows a reduction of $\sim$40\% (Fig. \ref{fig:nw_kthermal} d). The $\sigma_{l}$ reduction shows the following order $[111] \; \approx \;[$100$] \; > [$110$]$ for all the wire cross-section sizes considered here. Thus, $\sigma_{l}$ can be tuned by three ways, (i) wire cross-section size, (ii) hole radius and (iii) channel orientation.   

\textit{Reasons for the reduction in $\sigma_{l}$:} The observed trends in the reduction of $\sigma_{l}$ can be attributed to two reasons, (i) increased phonon localization in hollow nanowires and (ii) anisotropic modes and mode reduction in nanowires. As the nanowire size reduces the geometrical confinement increases which results in increased phonon confinement \cite{li_hollow_sinw}. All the phonon modes in hollow nanowires do not propagate well and become localized which reduces $\sigma_{l}$ compared to the solid Si and Ge NWs \cite{li_hollow_sinw,PR_paper}.  

\begin{figure}[b]
	\centering
	\includegraphics[width=2.6in,height=1.7in]{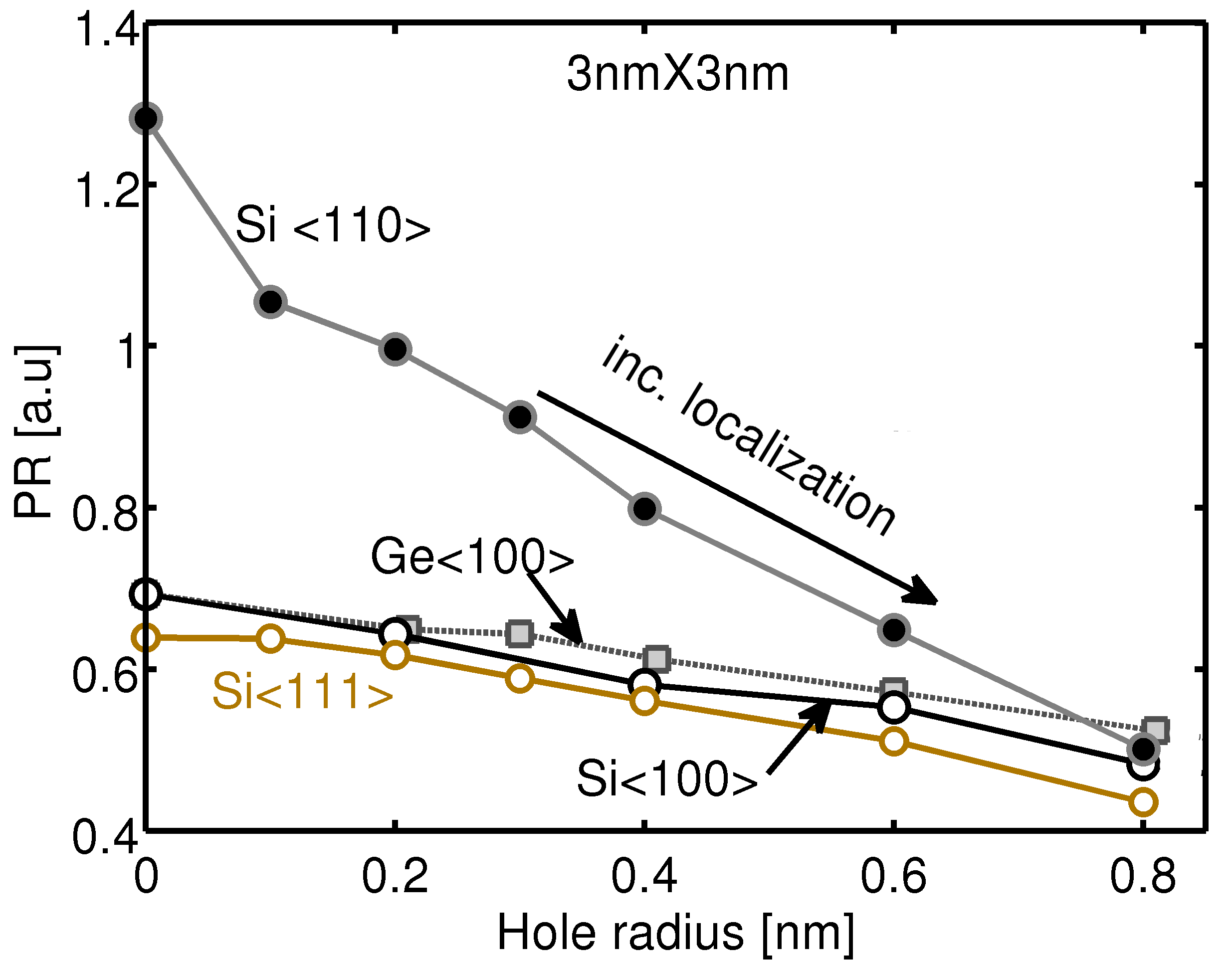}	
	\caption{ Average participation ratio (PR) in Si and Ge NWs. PR reduces with increasing hole radius indicating increasing phonon localization in hollow NWs. }
	\label{fig:energy_localization}
\end{figure}

\textit{(a) Phonon localization:} The extent of localization of phonon modes can be calculated using a `participation ratio' (PR) \cite{PR_paper}. This ratio can be calculated for each phonon mode as \cite{PR_paper}, 
\begin{footnotesize}
\begin{equation}
	\label{pr_eq}
	PR^{-1} = N\Sigma_{i}(\Sigma_{n,j\in[x,y,z]} \psi^{*}_{i,n,j} \psi_{i,n,j})^{2},
\end{equation}
\end{footnotesize}
where N is the total number of atoms in the unitcell, n and j represent the number of sub-bands and directional vectors, respectively and $\psi_{i,n,j}$ is the eigen vector associated with atom `i', sub-band `n' and direction `j'. The eigen vectors are calculated from the MVFF dynamical matrix formulation as described in Ref.\cite{vff_own_paper}. The participation ratio measures the fraction of atoms participating in a mode and hence varies between $O(1)$ for delocalized states to $O(1/N)$ for localized states and effectively indicates the fraction of atoms participating in a given mode.

\begin{figure}[t!]
	\centering
		\includegraphics[width=3.2in,height=2.1in]{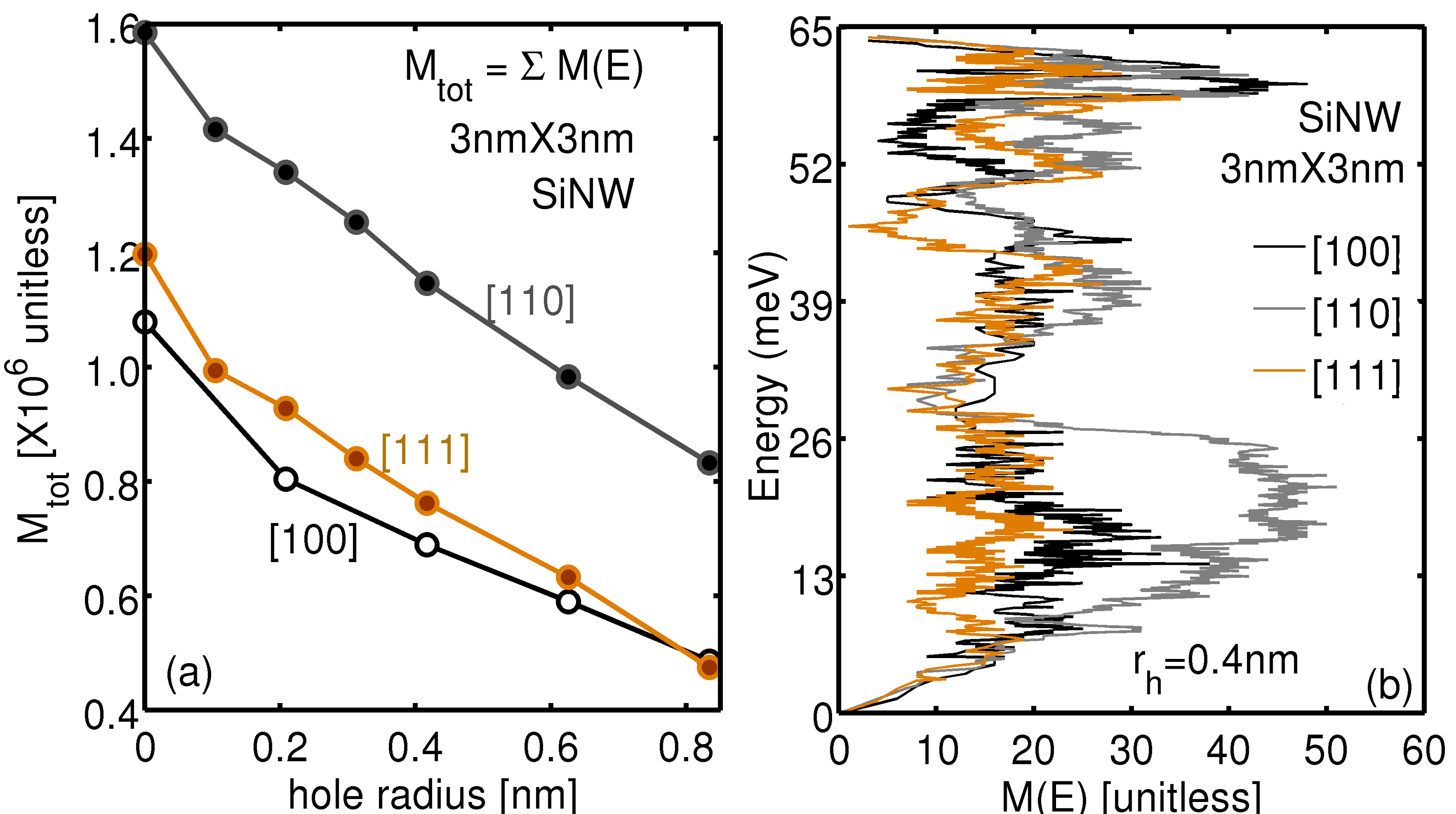}
	\caption{(a) Total modes in hollow SiNWs for three channel orientations with hole radius. $[$110$]$ has maximum modes. $[$111$]$ and $[$100$]$ are very close.(b) Modes distribution in energy for three channel orientations. Both plots are for 3nm $\times$ 3nm SiNWs.}
	\label{fig:modes_sinw}
\end{figure}

The average PR in 3nm $\times$ 3nm, $[$100$]$ Si and Ge NW shows a reduction with increasing $r_{h}$ (Fig. \ref{fig:energy_localization}). The solid NWs have PR close to 0.7, showing delocalization, which gradually decreases indicating phonon localization. Thus, both Si and Ge show similar phonon localization and hence similar $\sigma_{l}$ reduction (weak material dependence) (Fig. \ref{fig:nw_kthermal}a and b). 

\textit{(b) Anisotropic phonon modes:} The reduction in $\sigma_{l}$ with channel orientation shows anisotropy due to different propagating modes (M(E)) for each orientation. Figure \ref{fig:modes_sinw} (a) shows that $[$110$]$ wires have highest modes while $[$100$]$ and $[$111$]$ wires have fewer modes \cite{jauho_method}. $[$100$]$ and $[$111$]$ wires show similar decrease in total M(E) with $r_{h}$ (Fig. \ref{fig:modes_sinw} a) which explains the similar reduction in $\sigma_{l}$. The energy resolved M(E) for 3nm $\times$ 3nm SiNWs with $r_{h}$ = 0.4nm (Fig. \ref{fig:modes_sinw} b) clearly shows higher number of modes in $[$110$]$ orientation thus resulting in smaller reduction in $\sigma_{l}$. Phonon localization also reflects that a $[$111$]$ SiNW shows similar localization as $[$100$]$ SiNW whereas, $[$110$]$ SiNW shows less localization (Fig. \ref{fig:energy_localization}). This further corroborates the anisotropic reduction of $\sigma_{l}$ in SiNWs.

\section{Conclusion}
\label{sec:conc}

It has been shown that the presence of holes in Si and Ge NWs can be used for tuning their thermal conductance. Increased phonon confinement, phonon localization due to increased surface-to-volume ratio and mode reduction are the reasons for such drastic reduction in $\sigma_{l}$. Thus, variation of nanowire cross-section size, hole radius and channel orientation provide attractive ways to tune the thermal conductance. $[$100$]$ and $[$111$]$ nanowires show maximum reduction in $\sigma_{l}$ for Si. Similar trends are also expected in GeNWs. This can pave the way to make better TE devices using these nanowires.

\section{acknowledgment}
The authors acknowledge financial support from MSD, FCRP, MIND/NRI, and NSF and computational support from nanoHUB.org, an NCN operated and NSF funded project.


%

\end{document}